\documentclass[conference]{IEEEtran}

{}
{}
{}

\usepackage{subfig}
\usepackage{graphicx}
\usepackage{booktabs}
\usepackage[ruled]{algorithm2e}
\SetKwInput{KwData}{Input}
\SetKwInput{KwResult}{Output}

\usepackage{cite}
\usepackage{amsmath,amssymb,amsfonts}
\usepackage{textcomp}
\usepackage{xcolor}
\usepackage{gensymb}
\usepackage{tabularx,booktabs}
\usepackage{lipsum}

\begin{document}

\title{Reconfigurable Liquid Crystal Reflectarray Metasurface for THz Communications}



\author{\IEEEauthorblockN{Xiaomin Meng\IEEEauthorrefmark{1},
Maziar Nekovee\IEEEauthorrefmark{2} and Dehao Wu\IEEEauthorrefmark{3}}
\IEEEauthorblockA{\IEEEauthorrefmark{1}\IEEEauthorrefmark{2}\IEEEauthorrefmark{3}Department of Engineering and Informatics, University of Sussex, United Kingdom. \IEEEauthorrefmark{2}Quantrom Technologies Ltd.\\
Email: \IEEEauthorrefmark{1}xm51@sussex.ac.uk,
\IEEEauthorrefmark{2}m.nekovee@sussex.ac.uk,
\IEEEauthorrefmark{3}dehao.wu@sussex.ac.uk,
}
}
\maketitle


\begin{abstract}
We present computational studies on a proposed 20 by 20 elements electronically reconfigurable liquid crystal (LC) based binary phase reflectarray metasurface, operational at 108 GHz. LC was modelled after Mer's GT3-23001, and full wave simulations have shown a phase difference of 177 degrees between ON and OFF states, while reflection amplitudes were both 0.88 for ON and OFF. We present preliminary full wave simulation results on the Genetic Algorithm (GA) optimised far-fields. We also present the basic design procedures and cross-platform implementations on optimisation routines involving Matlab, CST and VBA environments.
\end{abstract}

\section{Introduction}\label{sec1}
\subsection{Motivation and Challenges}

In the field of wireless communication, as a result of the ever-increasing demands for faster link speeds and accompanying technological advancements, we have entered the era of milimeter wave and began exploration of the sub-millimetre world, as FAA announced the opening of 95 GHz to 3 THz spectrum for experimental 6G purposes \cite{1}. As we step higher into the frequency regimes, electromagnetic (EM) propagation become increasingly analogous to visible light propagation - this brings many new challenges to establishing stable and fast links.

One of the major issues with millimetre and sub-millimetre wave propagation is the non-line-of-sight (NLOS) problem. In short, the coverage of THz propagation will be greatly limited as propagation will be barricaded by most visible objects. More over, reflection, which is often utilised effectively in the channel modelling of lower frequency regimes, will also be vastly different in THz regime - many previously flat surfaces, such as concretes and bricks, will now appear bumpy to THz waves and hence disperse rather than reflect. Thus, THz propagation will ideally need to achieve direct link among transceivers.     

Another issue with achieving high speed link in high frequency regime is the high free-space-path-loss associated. Going from 1 GHz to 100 GHz is accompanied by 10,000 times increase in path loss, which immensely raises the gain requirements for antennas. In fact, if we were to adopt traditional phased array, we would roughly need to increase from 1024 elements in 28 GHz to 100,000 elements at 300 GHz, to maintain a stable link of around 20 Gbps over 200 meters. This exponential increase in phased array elements proves to be a challenge both technologically and practically, because firstly, power amplifier requirement for 100,000 element array will be extremely high, secondly, phase delay components at THz is extremely difficult to achieve, and last but not least, the losses associated with feeding 100,000 elements will make the device extremely inefficient.


In this paper, we propose the design and application of reconfigurable reflectarray metasurface in tackling above challenges in THz wireless communication. The liquid crystal based reconfigurable reflectarray is semi-passive, in the sense that power is only required to control the ON/OFF state of the individual antenna element, but not required to generate radiation, since the device will be reflecting an impinging source of EM wave. In this way the energy efficiency can be greatly improved; we envision a powerful source, such as an optical one, used to feed the device surface, while the device is acting as a ``smart-mirror'', reflecting the impinging source of EM propagation according to input configurations.

\subsection{LC based Reconfigurable Reflectarray Metasurfaces}

Traditional reflectarrays tend not to be reconfigurable, this is because their design rely on intrinsic properties of unit element antenna dimensions - by adjusting the dimensions of each antenna, a phase profile can be achieved for certain desired wavefront in the far-field. With advancements in the field, electronic reconfigurability was introduced through methods such as the incorporation of diodes\cite{2,3,4}, which provided an equivalence of 2 or more antenna dimensions through the controlling of ON and OFF state of the diode. This brought much attention to the field as the benefit of electronic reconfigurability is clear over the mechanical counterparts, such as using MEMS, which can be both costly and less reliable in harsher operational environments. 


LC based reconfigurable reflectarray metasurfaces have also been increasingly studied\cite{13,14,17}, due to their advantages in higher frequency regime. LC based unit antenna elements are structurally simpler to the diode counterpart, and also more suitable for high frequency applications, as PIN diodes can be lossy at THz regimes. The principle of LC based device relies on the nematic nature of LC molecules, which can be realigned according to electric field; with applied electric electric field (such as through capacitor plates), one can change the electric permittivity of the LC, which will modify the resonance of the antenna and result in a phase shift between the ON and the OFF state.

\subsection{Contribution and Paper Overview}
Many of the PIN diode based reconfigurable reflectarray devices are binary phase controlled (either ON or OFF), however these are mostly proposed for the sub-THz regime; many of the LC based reconfigurable reflectarray are continuous phase controlled (through continuous voltage), hence not as simple and energy efficient as binary phase based devices. We propose a binary phase LC based device that combines the benefits of simplicity in binary phase control and efficiency of LC in high frequency regime.

In this paper we also present in depth the simulation routine for designing reconfigurable metasurfaces. The routine involves three softwares/environments: CST, Matlab and VBA. CST is a computational EM solver, which we used for the full wave simulations, Matlab is well known softwares for engineers, we have used it for the optimisation of antenna design and theoretical far-field, as well as GA pattern synthesis, and VBA is a the language used in initialising full device model for CST simulation. 

\section{Problem Setup and Formulations}

\subsection{Unit Cell}\label{sec2.2}

\begin{figure}[h]
\centering
{\includegraphics[scale=0.35]{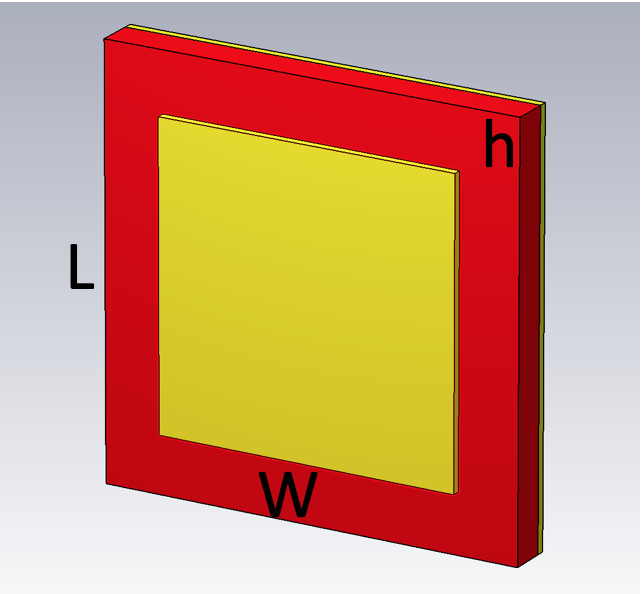}}
\caption{The schematics of unit cell design. $L$=1 mm, $W$=0.714 mm, $h$=0.087 mm.}
\label{uni}
\end{figure}

The LC we modelled after was the GT3-23001 that is commercially available from Merk. This LC has relatively stable properties of the permittivities across wide range of frequencies, as observed from different sources\cite{9,10,11,12,17}. Additionally, the operating temperature is -20$^{\circ}$\textasciitilde100$^{\circ}$, making outdoor application also possible. 

\begin{table}[h]
\centering
\begin{tabular}{|l|l|l|l|l|}
\hline
          & $\varepsilon_{\perp}$ & $\varepsilon_{\parallel}$ & $\tan \delta_{\perp}$ & $\tan \delta_{\parallel}$ \\ \hline
GT3-23001 & 2.47                 & 3.25    & 0.02    & 0.015    \\ \hline
\end{tabular}
\end{table}

The unit antenna element dimensions are shown in FIG-\ref{uni}, where we have optimised the parameters using a cross-platform routine, which we will present in the following section. The operational frequency of the device is 108 GHz, and the full device consists of 20 by 20 unit elements. The substrate, indicated as red coloured material is the LC, has parameters listed in the table. We simulate the ON and OFF configuration of the unit element by the permittivity and loss tangent values of the LC at those states. It is important to note that we have not included in the simulation the active controls of biasing circuit, in order to save computational time without significant impact on the accuracy of the results.

Although our unit element periodicity is only 1 mm, which means the dimension of our full device is 2 cm by 2 cm, for future work and possible application scenarios we intend to address cases of multiple 20 by 20 devices working collaboratively, or even single device with much greater number of elements, to increase the aperture area. Our optimism in a small aperture area also lies in the vision to adopt powerful optical sources that are much better at confining EM propagation in specific direction (laser like feed sources).

\subsection{Theoretical Far-field}\label{sec2.3}

\begin{figure}[h]
\centering
{\includegraphics[scale=0.28]{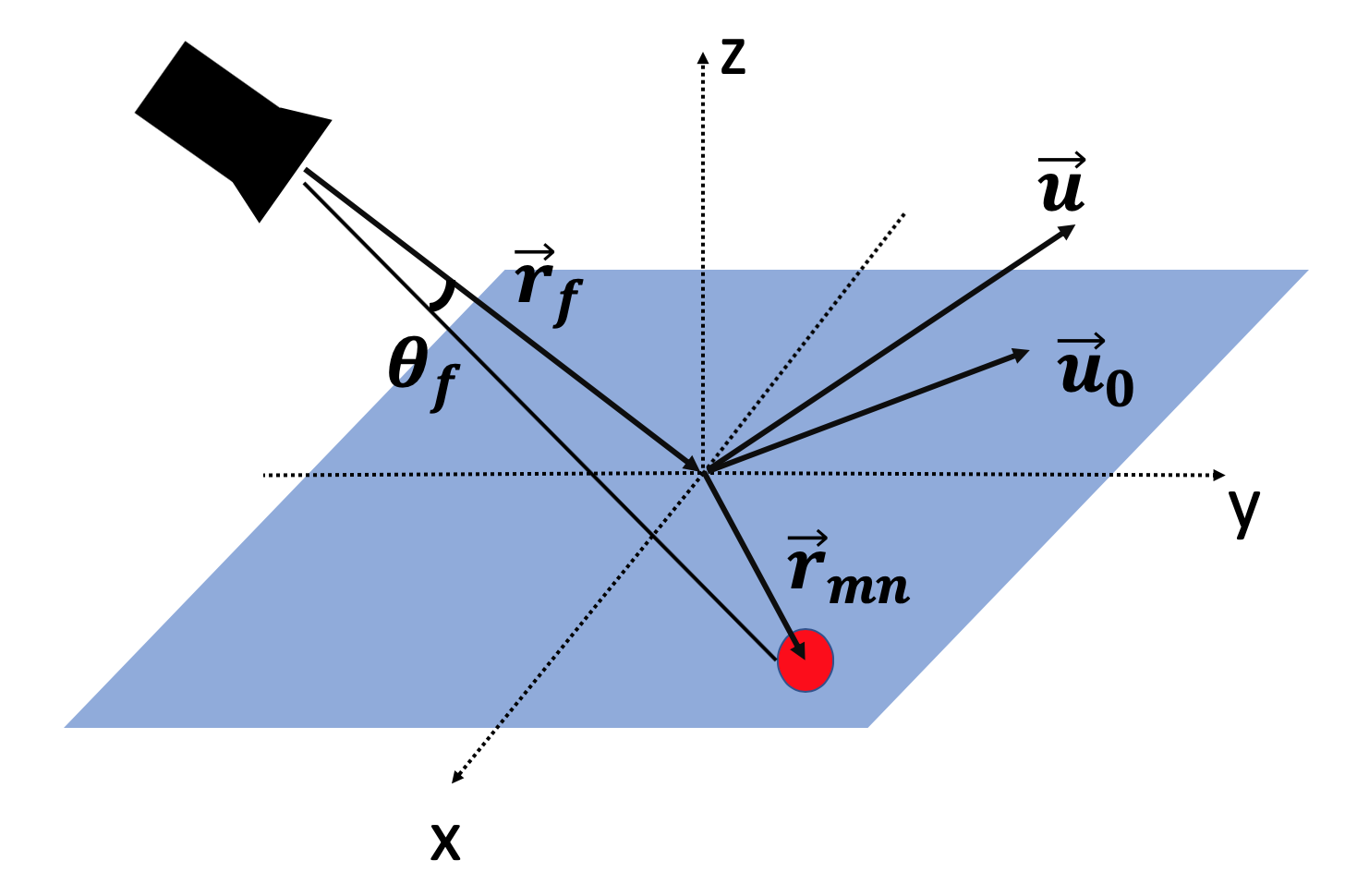}}
\caption{Geometrical conventions used for the theoretical far-field calculations.}
\label{illu}
\end{figure}

The radiation pattern, $E(\theta,\phi)$, is a product of unit patch antenna and feed horn radiation patterns, which are approximated as cosine function raised to the power of $k$, the array factor (first exponential term), which is simply summing up the spherical wavefronts from each unit antenna to observation point, and the amplitude ($\Gamma_{mn}$) and phase ($e^{i\phi_{mn}}$) resulting from the state of each unit:

\begin{equation}
\begin{split}
&E(\theta,\phi)= \\
 &\sum_{m=1}^{M} \sum_{n=1}^{N} \cos^q{\theta} \frac{\cos^{q}{\theta_f}}{|\vec{r}_{mn}-\vec{r}_f|} \cdot e^{-ik(|\vec{r}_{mn}-\vec{r}_f |-\vec{r}_{mn} \cdot \hat{u})} \Gamma_{mn} \cdot e^{i\phi_{mn}}
\end{split}
\end{equation}

\begin{gather}
\phi_{mn} = 
\begin{bmatrix}
    1/0 & 1/0 & \dots  & 1/0 \\
    1/0 & 1/0 & \dots  & 1/0 \\
    \vdots & \vdots & \ddots & \vdots \\
    1/0 & 1/0 & \dots  & 1/0
\end{bmatrix}
\cdot \phi_{\Delta}\label{eq2}
\end{gather}

\subsection{Genetic Algorithm}\label{sec2.4}
To achieve multi-functionality in terms of beam manipulation, we have implemented a genetic algorithm (GA) in order to optimise the binary matrix.

\begin{figure}[htb]
\centering
{\includegraphics[scale=0.48]{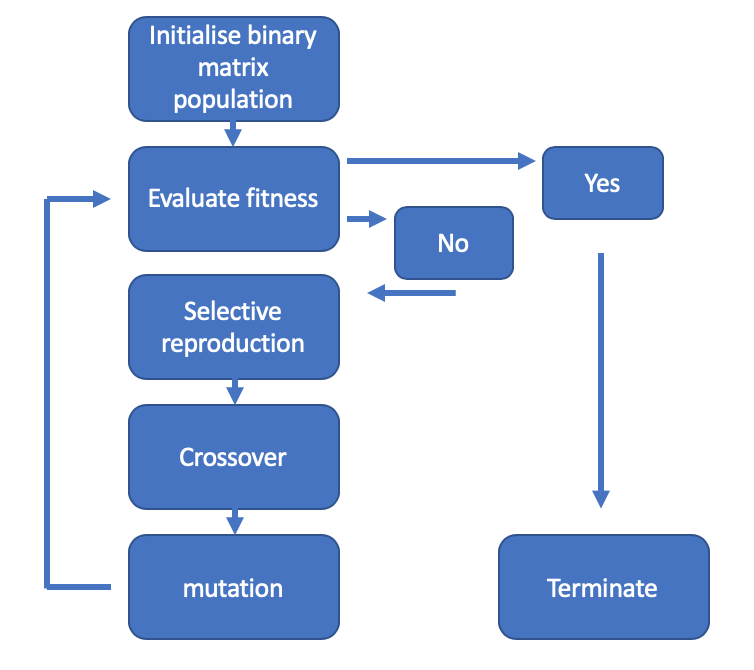}}
\caption{GA algorithm flowchart\label{fig7}}
\end{figure}

For the initial population, we have a codebook for known common beam-functionalities, such as the checkerboard and the stripes configurations of the binary matrix. The cost function that determines the fitness of our far-field results is defined as follows:

\begin{equation}  cost = |E - E_{target}|^2  \end{equation}

where $E_{target}$ is the desired far-field electric field pattern that we wish to find out about the coding matrix. More details on the cost function are presented in the Implementation section.

\section{Implementation Rountines}
\subsection{Unit Cell}
The unit cell design was achieved in a few stages: We first adopt a preliminary design of the dimensions using formulas of patch antenna design, confining the size of the patch to be around half the wave length. The periodicity was arrived at considering the balance between beam resolution and steering angle range. As a side step we check the beam profiles of various configurations with theoretical far-field plot.

\begin{figure}[htb]
\centering
{\includegraphics[scale=0.48]{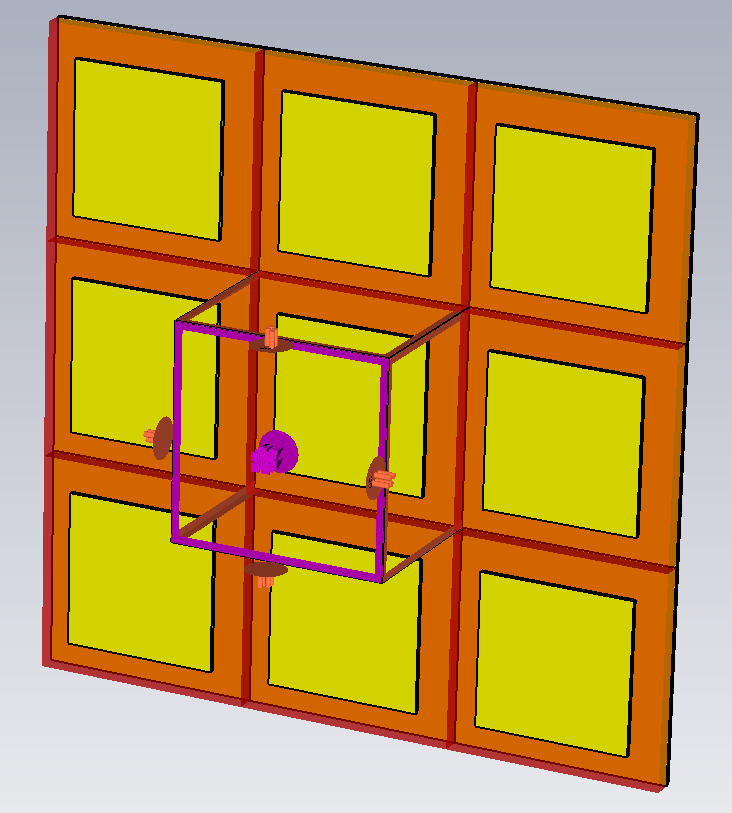}}
\caption{The unit cell simulation with periodic boundary condition.}
\end{figure}

When the preliminary parameters are set, we develop the unit cell model in CST, for which we have set the unit cell boundary conditions at the plane parallel to device surface, to mimic an infinite system. The incident port set normal to the device surface, in which case the S11 parameter indicates the reflection parameter away from surface. This is to be contrasted with traditional patch antenna designs, where the antenna is usually fed through a feed-line, for which the S11 parameter will indicated the reflection through feed-line, which is normally a plane perpendicular to patch surface. 

With these parameters in place, we perform a parameter sweep (in our case on the height $h$ and the patch width $W$) over range of values, and collect the amplitude and phase of the S11 parameter. Performing algorithm-1, we will arrive at the optimal unit cell structure parameters.

\begin{algorithm}[h]
\caption{Matlab script to optimise unit cell dimensions}
 \KwData{S11p - S11 phase; S11a - S11 amplitude; hu, hl - LC height lower/upper bound; Wl, Wu - patch antenna width lower/upper bound; pl, pu - phase difference lower/upper bound; al, au, adm - amplitude lower/upper bound and amplitude difference margin}
 \KwResult{h, W - Unit cell height and width optimised for close to 180 degrees phase difference, and maximal reflection amplitude (minimal difference) of both states}
 initialise pu, pl, au, al\;
 \For{h=hl $\rightarrow$ hu}{\For{W=Wl $\rightarrow$ Wu}{
S11P\{h,W\}=S11p$_{on}$\{h,W\}-S11p$_{off}$\{h,W\}\;
S11A\{h,W\}=S11a$_{on}$\{h,W\}-S11a$_{off}$\{h,W\}\; }
 }
Result $\leftarrow$ Find (S11a $\geq$ al) and (S11a $\leq$ au) and (S11A $\leq$ adm) and (S11P $\geq$ pl) and (S11P $\leq$ pu)\;
Result = Sort Result\;
h, W $\leftarrow$ Result
\end{algorithm}

The optimisation of the unit cell is then finalised by exporting the S11 results into Matlab and running our routine. Our routine will first compute the phase and amplitude differences, then select the optimised parameters given the condition of 1) phase difference of close to 180 degrees, 2) minimal amplitude difference, 3) maximal absolute amplitudes

\subsection{Theoretical Far-field and GA}
The theoretical far-field is implemented in Matlab, for which we can obtain the radiation pattern according to specific configuration binary matrix input. For pattern synthesis, where we address the inverse problem of knowing a specific desired radiation pattern and solving for the configuration binary matrix, we have chosen to adopt GA optimisation for a numeric approach.

There is a GA optimisation toolbox available in Matlab, however we have also implemented our own GA algorithm for more flexibility in the operations performed. The general implementation is outlined here:

\begin{algorithm}[h]
\caption{Matlab script for GA optimisation on beam-steering pattern synthesis}
 \KwData{LOC - index location of desired peak}
 \KwResult{Matrix\textunderscore f - final configuration/population}
Initialise Matrix with random 1's and 0's\;
\While{(Counter $\leq$ max generation) or (cost $\leq$ max accepted)}{
Compute E with Matrix\;
LOC\textunderscore E $\leftarrow$ peak index of E\;
cost $\leftarrow$  $|$LOC - LOC\textunderscore E$|$$^2$\;
----------------------------------------------------------------\\
Matrix\textunderscore new $\leftarrow$ crossover, mutation, inversion on Matrix\;
----------------------------------------------------------------\\
compute E\textunderscore new with Matrix\textunderscore new\;
LOC\textunderscore E\textunderscore new $\leftarrow$ peak index of E\textunderscore new\;
cost\textunderscore new $\leftarrow$ $|$LOC - LOC\textunderscore E\textunderscore new$|$$^2$\;
----------------------------------------------------------------\\
\If{cost\textunderscore new $<$ cost}{update Matrix with Matrix\textunderscore new}
}
\end{algorithm}

The cost function for our GA is defied as the modulus squared of the target beam profile and desired beam profile. Specifically, if we were interested in achieving beam-steering, we first find the index location of the peaks of the far-field beam profile, then the cost function actually becomes 

\begin{equation}  cost = |E - E_{pk1}|^{k} + |E - E_{pk2}|^{k-1} +|E - E_{pk3}|^{k-2} + ...  \end{equation}

where $E$ is the index of the desired max electric field, $E_{pk}$ is the index of the electric field strength at a specific peak, where electric field strength is highest at pk1, $k$ is the exponential of importance. We assign the cost of highest peak to be of most importance, thus giving it highest exponential, to allow the algorithm to quickly optimise for beam-steering at desired location. However, we also impose secondary cost conditions, enforcing that the sidelobes are either the mainlobe, or shall have minimal amplitude (essentially sidelobe level reduction).

Similar methods to achieve other beam-profiles can be attained with the same logic. For instance when we optimised for multi-beam profile, we simply designed the cost function as:

\begin{equation}  cost = |E_1 - E_{pk1}|^{k} + |E_2 - E_{pk2}|^{k} +|E_3 - E_{pk3}|^{k} + ...  \end{equation}

where the modification has been that the desired max electric field has been replaced with $E_1, E_2, E_3$, which are the electric field index at desired peak 1, peak 2 and peak 3 field strengths. These terms are assigned with equivalent importance exponent, while also emphasising the sidelobe reduction with the following terms. In short, the careful design of cost function is as important (if not more important) as the designing of GA evolution operations.

\subsection{Full Device}

When the configuration of ON/OFF states is known, we then perform full wave simulation in CST environment. However, the setting-up of the ON and OFF cannot be automated unless a VBA script is implemented to assist the CST program. In algorithm-3, we initialise a 20 by 20 full device with unassigned permittivity values, we then read line by line the input configuration of ON/OFF and assign the permittivity values according to their states:

\begin{algorithm}[h]
\caption{VBA script to initialise structure in CST according to configuration}
 \KwData{FILE - configuration of ON/OFF, 20 by 20 matrix}
 \KwResult{Full device model updated according to ON/OFF configuration and ready for full wave simulation}
 Initialise 20 by 20 structure with permittivity $\varepsilon$ undefined\;
 \While{not end of line}{
  A $\leftarrow$ read current line\;
  B $\leftarrow$ split(A) with delimiter\;
  \For{counter1, counter2}{
   Matrix(counter1,counter2)=B(counter1,counter2)\;
   update counters\;
   }
 }
 \For{i=1:20}{
 \For{j=1:20}{
 \eIf{Matrix(i,j)==1}{
 $\varepsilon_{i,j}$=2.46; $\tan \delta=0.02$;}{
 $\varepsilon_{i,j}$=3.28; $\tan \delta=0.015$;}
 }
 }
\end{algorithm}

\section{Results and Conclusion}

\subsection{Unit Cell and GA Optimisation}
In FIG-5, we have shown that at 108 GHz, the device can attain a phase difference close to 180$^{\circ}$ between the reflected wave in ON and OFF states, meanwhile the reflected amplitude is maintained at 0.88 for both ON and OFF reflection.
\begin{figure}[h]
    \centering
  \subfloat[\label{a}]{
        \includegraphics[width=0.95\linewidth]{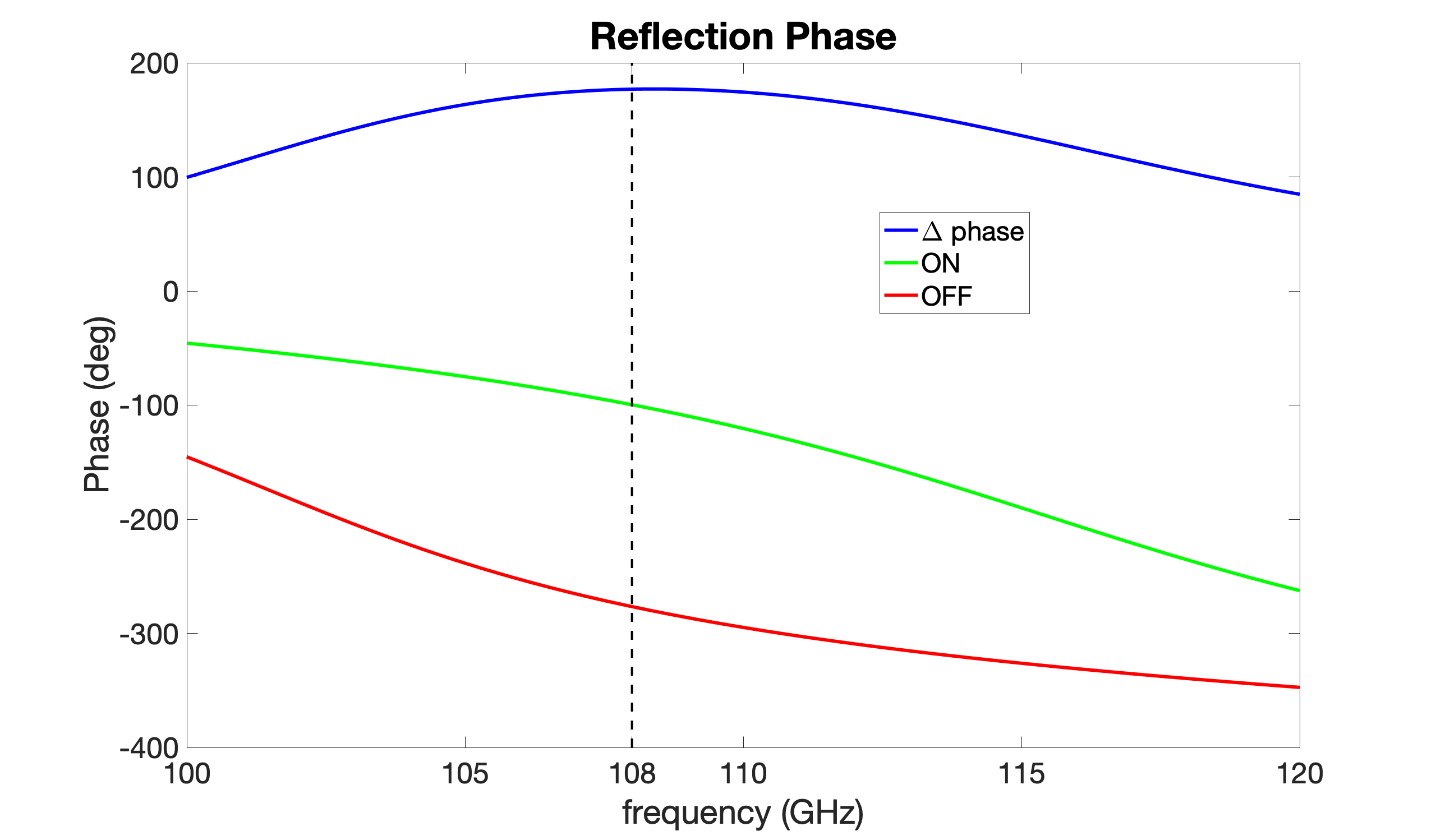}}
    \\
  \subfloat[\label{b}]{
        \includegraphics[width=0.95\linewidth]{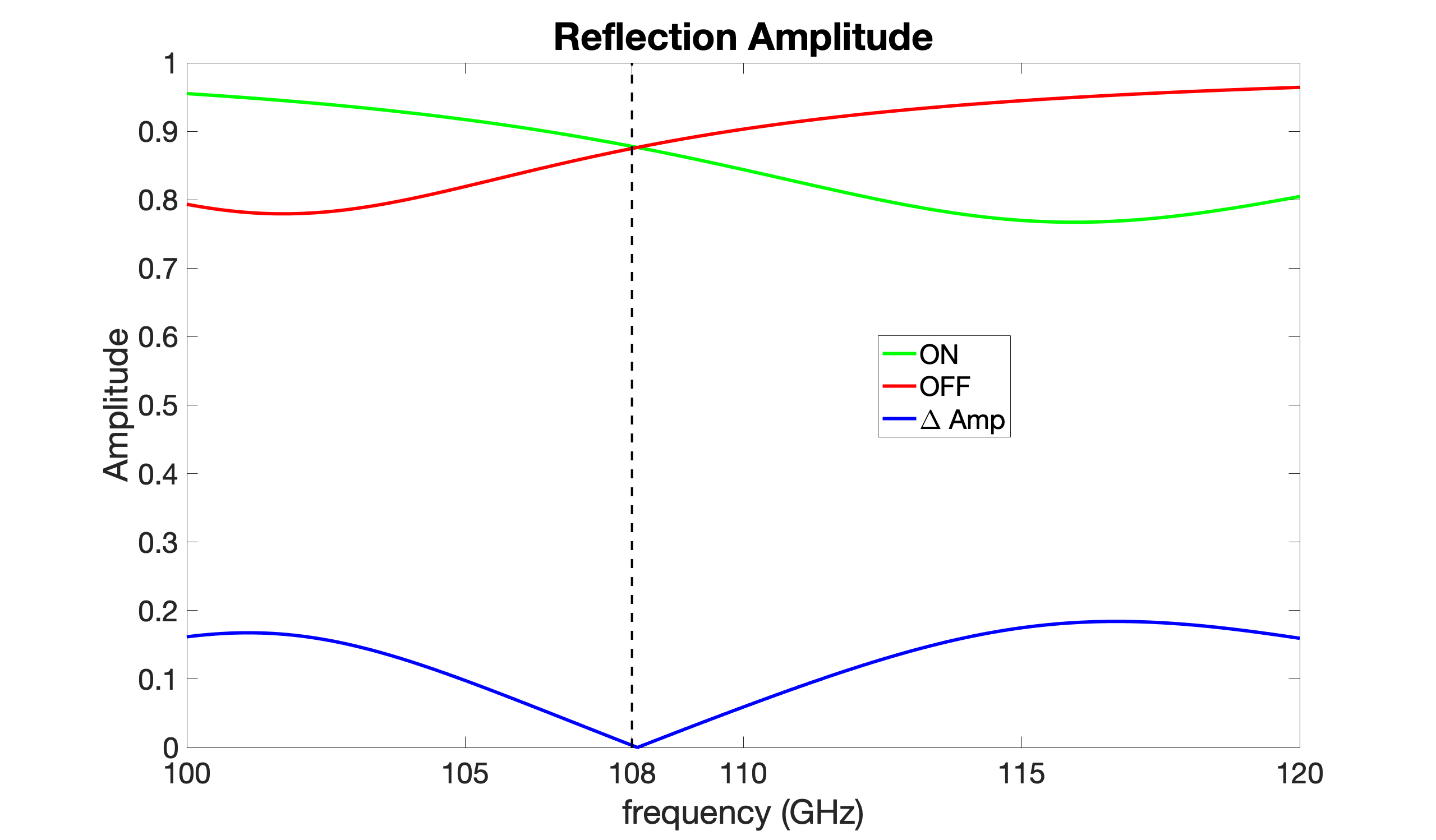}}
\caption{Unit cell reflection phase and amplitude properties.}
\end{figure}

In FIG-6, we present a set of results showing the GA optimisation for pattern synthesis. In FIG-6 a), we show the radiation pattern plot from a far-field plane wave source incident at 54$^{\circ}$ ($[-1;-1;1]$ direction), whereas in b), the same plane wave 54$^{\circ}$ offset source is received, but now the configuration matrix has been GA optimised to beam-steer towards $\theta=45^{\circ}$, $\phi=135^{\circ}$. Corresponding ON/OFF configurations are shown in the plots.

\begin{figure}[h]
    \centering
  \subfloat[\label{a}]{
        \includegraphics[width=0.5\linewidth]{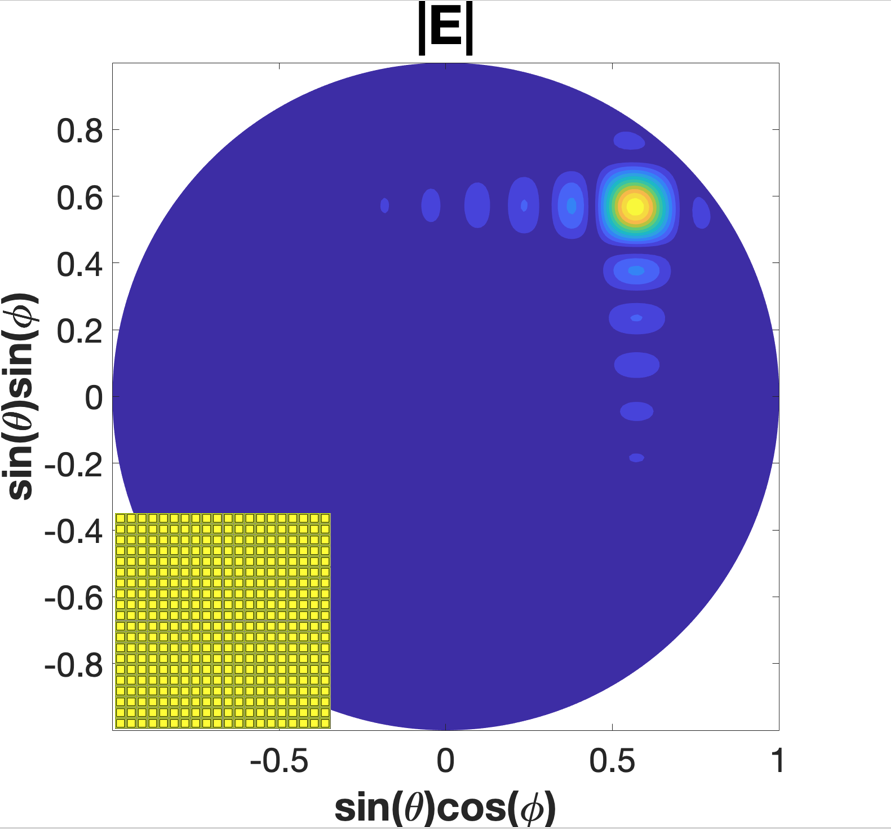}}
  \subfloat[\label{b}]{
        \includegraphics[width=0.5\linewidth]{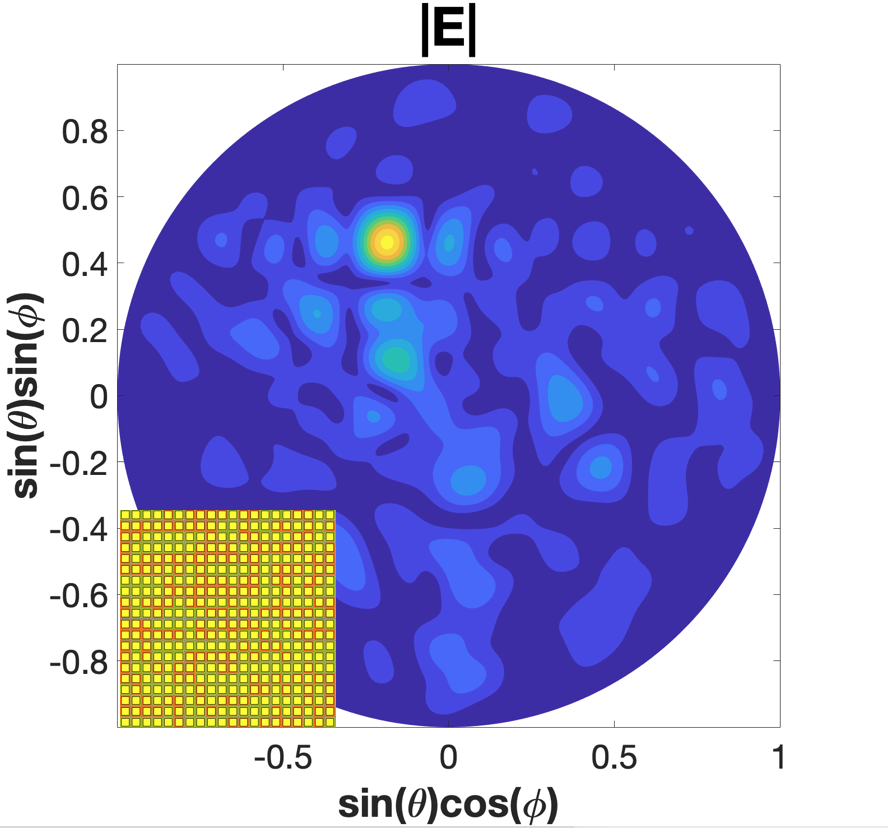}}
\caption{a) Radiation pattern plot from a far-field plane wave source incident at 54$^{\circ}$ ($[-1;-1;1]$ direction) b) same plane wave 54$^{\circ}$ offset source but now optimised to beam-steer towards $\theta=54^{\circ}$, $\phi=135^{\circ}$. Corresponding ON/OFF configurations are shown next to their plots.}
\end{figure}

\newpage
\subsection{Full Wave Results} 

In this section, we present 3 sets of results: 1) far-field plane wave normal incidence on the device, 2) $54^{\circ}$ offset far-field plane wave source, 3) near-field feed horn situated at $L \cdot [-4.5;0,13]$ with respect to the device's centre. 

\begin{figure}[htb]
    \centering
  \subfloat[]{
       \includegraphics[width=1\linewidth]{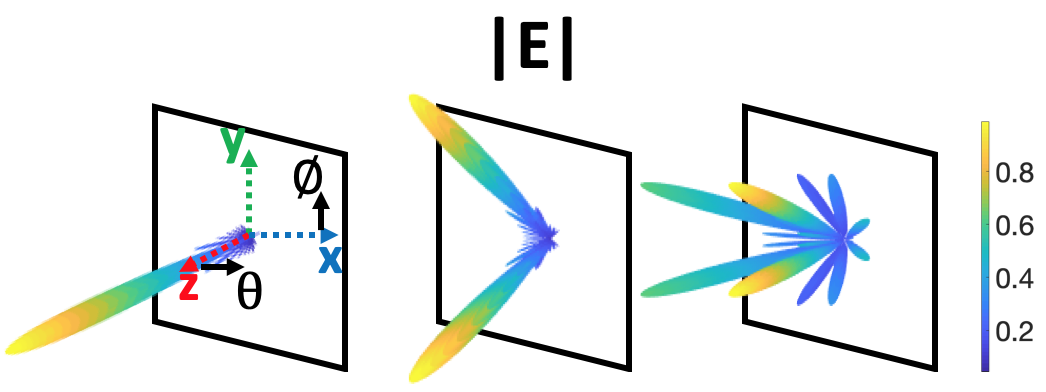}}
    \\
  \subfloat[]{
        \includegraphics[width=1\linewidth]{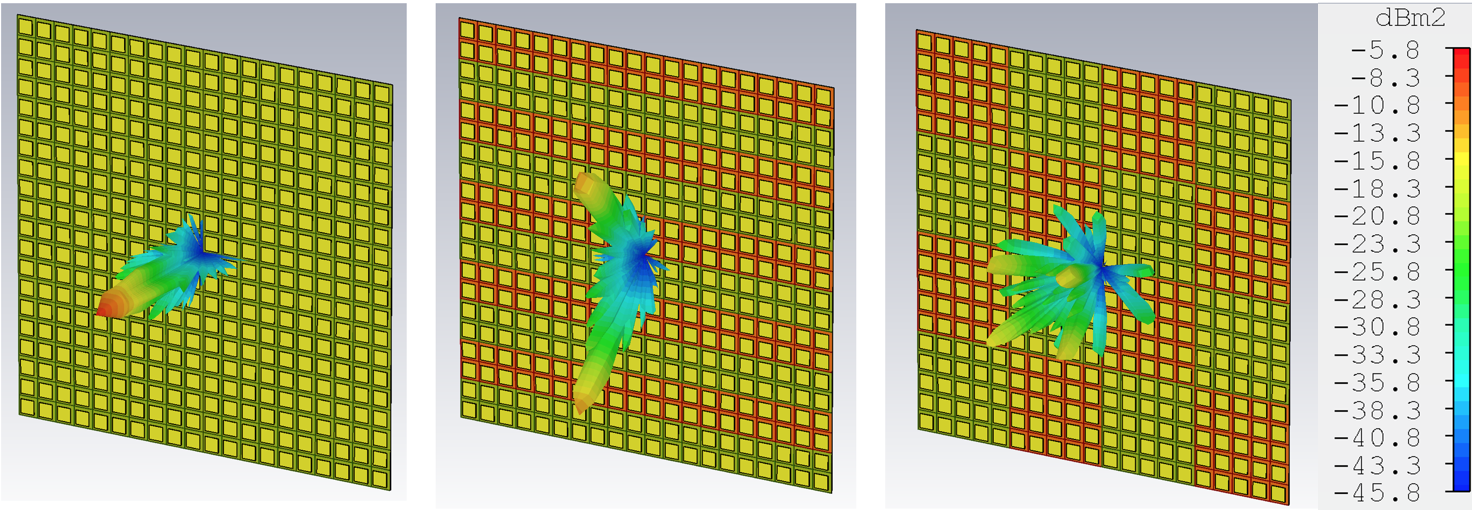}}
\caption{Green/red coloured unit element for ON/OFF state. a) theoretical far-fields for some canonical configurations, normalised to their maximum b) full-wave simulation of the same configurations. Note the area of the device is $20mm^2$, so at peak RCS of -5.8 dBm2, it is equivalent to linear RCS of $260mm^2$.}
\end{figure}

In FIG-7, we have simulated a linearly-polarised (along x) normal incident plane wave, and the results are qualitatively consistent with those of the PIN-diode based papers\cite{2,3}, given the same configuration. Note the use of RCS for units of far-field, as we simulate the scattering of an impinging plane wave, thus dBi units would be rather irrelevant due to the lack of a powered feed source. 
 
\begin{figure}[htb]
    \centering
  {
       \includegraphics[width=1\linewidth]{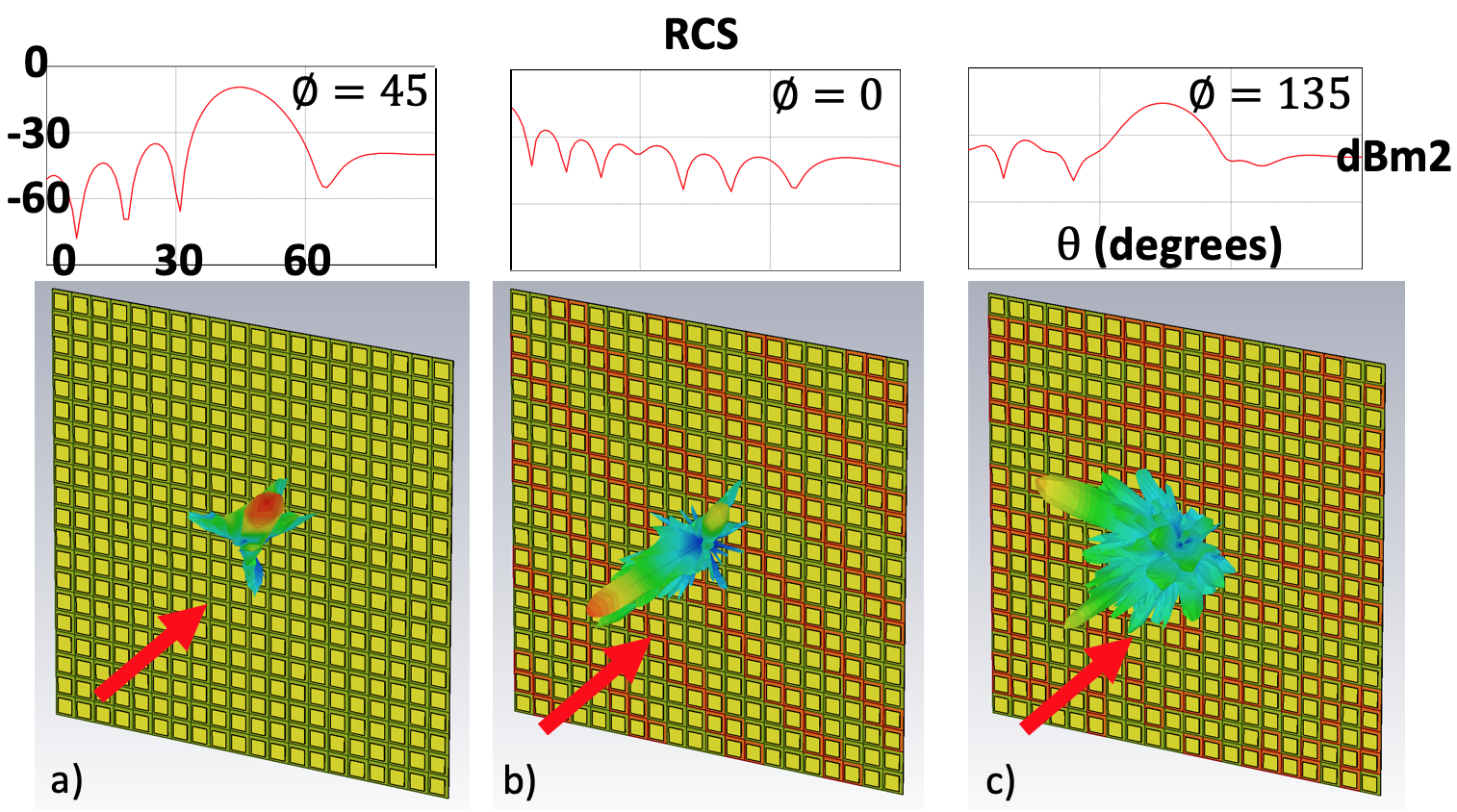}}
\caption{Full wave simulation of far-fields with 54$^{\circ}$ offset plane wave. a) all ON configuration, b) anomalous reflection profile and c) GA optimised beam-steering.}
\end{figure}

In FIG-8, we have simulated a 54$^{\circ}$ degrees incident plane wave from the direction of $[-1;-1;1]$. In FIG-8 a), when the configuration is all ON, we observe what is expected: mirror like reflection of the incident wave; in b), we proposed an anomalous reflection profile, where the beam is reflected towards the surface normal; in c), we have optimised the configuration matrix with GA to achieve beam-steering in the direction of $\theta=45^{\circ}$ and $\phi=135^{\circ}$.

\begin{figure}[htb]
    \centering
  {
       \includegraphics[width=1\linewidth]{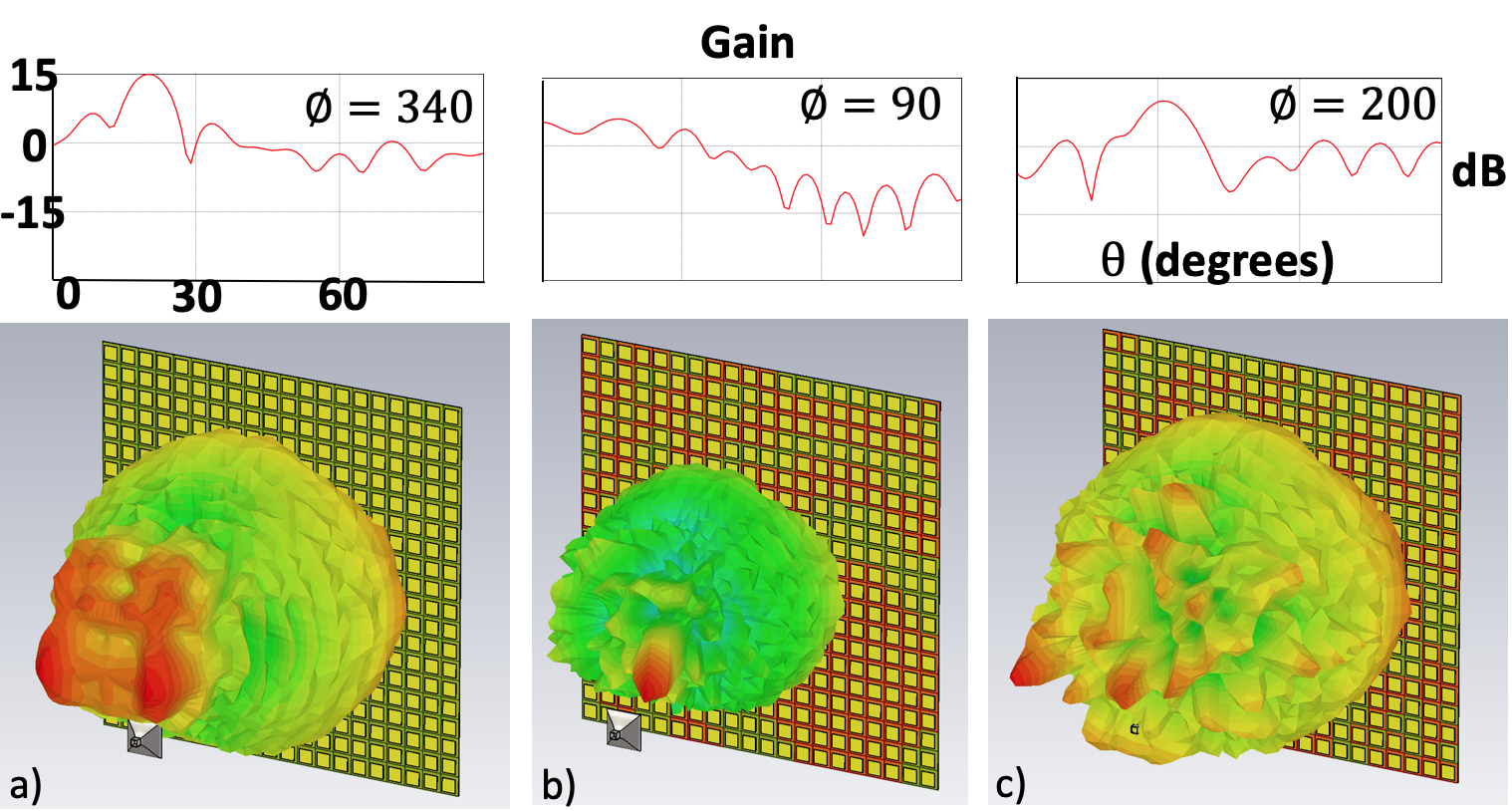}}
\caption{Full wave simulation of far-fields with close-up feed-horn. a) all ON configuration, b) proposed beam-steering, c) GA optimised multi-beam profile.}
\end{figure}

In FIG-9, we have simulated a feed horn located near the device at $L \cdot [-4.5;0,13]$, which is visible on the plots. In FIG-9 a), the configuration matrix is all ON, thus we observe as expected a reflection of the point source pattern; in b), the configuration for a focusing profile adapted to point source results in an off-set beam steering; and in c), our preliminary GA optimisation for multi-beam profile with mainlobe at the direction of $\theta=30^{\circ}$ and $\phi=200^{\circ}$. 

Note the units of far-field in dB, due to a realistically powered feed-source (rather than fictitious plane wave). The gain results can be improved as we have not optimised our feed-horn to provide efficient radiation coverage. Due to the early stage of our pattern synthesis work with GA, the multibeam profile has quite high sidelobe levels, which we are still working on reducing.

\subsection{Conclusion Remarks}

In this paper, we performed computational study on a proposed 20 by 20 element LC based binary phase reconfigurable reflectarray metasurface, which operates at the central frequency of 108 GHz. The phase difference between ON and OFF state achieved at central frequency is 177 degrees, and the reflection amplitudes are 0.88 for both states.

We presented some preliminary full-wave simulation results, for three cases: 1) normal incident plane wave, 2) 54 degrees incident plane wave, 3) offset near-field point source. We have also presented detailed implementation techniques employed to achieve the cross-platform simulation and optimisation. In terms of pattern synthesis, we showed the basic GA implementation and the principle ideas of designing the cost function for specific beam profile. 

For future work, we would like to improve the performance of GA pattern synthesis and address problems of the its computational efficiency, through both theoretical and hardware approaches. In terms of theory, we aim to adapt fast discrete Fourier transforms to save computational time, and in terms of hardware, we aim to implement parallel computing for the optimisation process. So far, the brute-force GA optimisation takes a few hours to complete an acceptable optimisation of 20 by 20 element pattern synthesis, which is way too slow for generating a pool of configurations for practical applications.

\section{Acknowledgements}
We would like to thank Dr. Richard Rudd from Plum Constulting, Dr. Junaid Syed from Rosenberger and Professor Marco Peccianti for the helpful discussions on this work.


%


\end{document}